
\magnification=1200
\hsize=15truecm
\vsize=23truecm
\baselineskip 18 truept
\voffset=-0.5 truecm
\parindent=1cm
\overfullrule=0pt

\centerline
{\bf Twistor--like Formulation of D=10, type IIA,
Superstrings $^*$}

\vskip 2truecm

\centerline{\bf P. Pasti and M. Tonin}

\vskip 1truecm

\centerline
{\sl Dipartimento di Fisica ``G. Galilei" -- Universit\`a  di Padova}
\vskip 1truecm
\centerline
{\sl Istituto Nazionale di Fisica Nucleare - Sezione di Padova}
\vskip 0.5truecm
\centerline{\sl Padova (Italy)}

\vskip 2truecm

\noindent
{\bf Abstract.} The twistor--like formulation of the type IIA superstring
$\sigma$--model in D=10 is obtained by performing a dimensional reduction
of the recently proposed twistor--like action of the supermembrane in
D=11. The superstring action is invariant under local, worldsheet $(n,n)$
supersymmetry where $3\leq n \leq 8$ and is classical equivalent to the
standard Green--Schwarz action (at least for $n=8)$

\vskip 2truecm

\noindent
$^{(*)}$ Supported in part by M.P.I. This work is carried out in the
framework of the European Community Research Programme ``Gauge Theories,
applied supersymmetry and quantum gravity" with a financial contribution
under contract SC1-CT92-D789.

\vskip 1truecm

\noindent
{\bf DFPD/94/TH/05}\hfill {\bf January 1994}

\vfill\eject

\noindent
{\bf 1 -- INTRODUCTION}

\vskip 0.5truecm

Recently some attention has been devoted to new formulations
$^{[1]-[14]}$ of superparticles, Green--Schwarz superstrings
and, in general, super--branes that involve twistor--like
$^{[15]}$ or harmonic variables $^{[16]}$. The motivation
of these attempts is the hope to clarify the role of the
$\kappa$--symmetry in these models and to have some insight
on the problem of their covariant quantization.

In particular, the twistor--like approach, with manifest
world manifold supersymmetry, first proposed by Sorokin et
al. $^{[1]}$ for superparticles in D=3,4, has been worked
successfully for $\sigma$--models of superparticles $^{[6]}$,
heterotic strings $^{[4],[5],[7],[9]}$ in D=3,4,6,10 and
supermembranes $^{[12]}$ in D=11.
Moreover type II superstrings have been considered by
Chikalov and Pashnev  $^{[10]}$ in D=4 with (1,0) world sheet
supersymmetry and by Galperin and Sokatchev $^{[11]}$  in D=3
with (1,1) world sheet supersymmetry. Recently a formulation of heterotic
string with both Virasoro constraints solved in twistor form has been
presented $^{[13]}$.
The approach of ref. [12]
has been also extended to other $p$--branes by Bergshoeff and
Sezgin $^{[14]}$.

The target space background of these $\sigma$--models fulfils
the same constraints that implement the $\kappa$--symmetry in the
standard formulations. In particular, for superparticles, D=10 heterotic
strings and D=11 supermembranes, these background constraints force the
SYM and/or SUGRA background to be on shell.

Moreover the embedding of the d--dimensional world supermanifold into
the D--dimensional target superspace is restricted by the ``twistor
constraint": the components of the pull--back of the vector supervielbeins
along the fermionic directions of the world supermanifold tangent space
vanish.

Thank to these background and twistor constraints, the background
two superform B which is present in these models, exhibits a
remarkable property, called Weyl triviality. This property is crucial
to get actions with a manifest, world manifold $n$--extended local
supersymmetry. This supersymmetry replaces $n$--components (and therefore
provides a geometrical meaning) of the $\kappa$--symmetry of the
standard formulations.

In this paper we pursue the program of the twistor--like approach by
presenting a twistor--like classical action for the D=10, type IIA,
superstrings $\sigma$--model.
Our action exhibits left--handed and right--handed,
$n$--extended, world--sheet local supersymmetry where $3\leq n\leq 8$
and is classically equivalent to the standard G.S. action.

For our purposes, twistors are just commuting Majorana spinors,
$\lambda$,
(Weyl--Majorana in D=10) in a space--time with Minkowski signature. Their
usefulness in twistor--like models is mainly due to the fact that in
D=3,4,6,10 dimensions, the vector $v^{\underline a}=\lambda\Gamma^{
\underline a}\lambda$ is automatically light--like. This follows from
the cyclic identity of the $\Gamma$--matrices in these dimensions. So
in the case
of heterotic strings, the twistor condition ${\cal E}_-^{\underline a}=
\lambda\Gamma^{\underline a}\lambda$ implies the Virasoro constraint
${\cal E}_-^{\underline a}{\cal E}_{-{\underline a}}=0$
(${\cal E}_\pm^{\underline a}$ are the left--handed and right--handed
components of the pullback of the target vector--like supervielbeins).
If one tries to repeat the same strategy for type II superstrings one
meets a difficulty. Indeed the twistor constraint in this case gives

$$
{\cal E}_-^{\underline a} = \lambda\Gamma^{\underline a}\lambda +
\bar\lambda \Gamma^{\underline a}\bar\lambda
$$

\noindent
and a similar relation for ${\cal E}_+^{\underline a}$. Here $\lambda$,
$\bar\lambda$ are two independent twistors (of opposite chirality for
type IIA superstrings) so that the Virasoro constraints for
${\cal E}_\pm^{\underline a}$ are not guaranted. In our model this
difficulty is overcome by the same mechanism at work for D=11
supermembranes. The point is that the worldsheet metric induced by the
target supervielbeins $E^{\underline a}$ does not coincide with the
metric specified by the preferred local frame where the $n$--extended
worldsheet supersymmetry  manifests itself. In other words the Virasoro
constraints do not appear in this local frame but in a different one.

The simplest way to get the correct constraints for the D=10,
twistor--like, type IIA, superstrings is to perform a dimensional
reduction of the D=11, supermembrane model of ref. [12], along the line
of ref. [17]. This is done in section 2. In section 3 we derive some
useful identities that follow from these constraints and we show that
the property of Weyl triviality is satisfied.
Finally in section 4 we write the action and we prove that it is
classically equivalent to the standard Green--Schwarz action.

Our notations are these of ref. [12]. In particular we shall follow
the convention of ref. [9] to use the same letters for the world
manifold and target space indices and to make the distinction by
underlining the target space ones. Moreover in order to distinguish
between D=11 and D=10 quantities and/or indices we shall put a
hat on the former.

An appendix collects our conventions about $\Gamma$ matrices in D=11
and their reduction in D=10 dimensions.

\vskip 0.5truecm

\noindent
{\bf 2 -- DIMENSIONAL REDUCTION}

\vskip 0.5truecm

The supermembrane superworld volume ${\cal M}(3|2n)$ is described by
the local coordinates

$$
\hat\zeta^{\hat{\cal M}}\equiv
(\xi^{\hat m}, \eta^{q\mu});\quad
\hat m=0,1,2;\ \mu=1,2;\ q=1,...n
$$

\noindent
where $\eta^{q\mu}$ are grassmann variables. We shall also use the
notation

$$
\xi^{\pm}=\xi^0\pm\xi^1\quad
\xi^2 = \rho
$$

The superstring superworldsheet ${\cal M}(2|n,n)$ is the slide of
${\cal M}(3|2n)$ at $\rho=0$. The dimensional reduction of the
worldmanifold is obtained by restricting all the superfields to be
independent from $\rho$ and by setting:

$$
\hat e^\pm = e^\pm(\zeta)\eqno(2.1a)
$$

$$
\hat e^{q\alpha} = e^{q\alpha}(\zeta) + {1\over 2}
\hat e^\perp\sigma^{\alpha\beta}_\perp D^q_\beta\psi(\zeta)
\eqno(2.1b)
$$

$$
\hat e^\perp = e^{-\psi(\zeta)} (d\rho + h(\zeta))\eqno(2.1c)
$$

\noindent
where $\zeta^M=(\xi^\pm, \eta^{q\mu})$, $\sigma^\perp_{\alpha\beta}$ is the
matrix ${0\ 1\choose 1\ 0}$ (and $\sigma_\perp^{\alpha\beta}$ is its
inverse) and $h=e^+h_+ + e^- h_- + e^{q\alpha} h_{q\alpha}$. The
one--superforms $\hat e^{\hat A}, (\hat A=\pm,\perp,q\alpha)$ and
$e^A, (A=\pm,q\alpha)$ are the d=3 and d=2 supervielbeins respectively.
Their duals are the vector superfields $\hat D_{\hat A}$ and $D_A$.
The d=2 torsion is

$$
{\cal T}^A = \Delta e^A = de^A + e^B \omega_B\ ^A
$$

\noindent
where $\omega_B\ ^A$ is a connection of the structure group
$SO(1,1)\otimes SO(n)$. With eqs. (2-1) and a suitable choice of the
reduction--redefinition of the connections, the d=3 constraints, as given
in [12], give rise to the standard, d=2, torsion constraints:

$$
{\cal T}^\pm = e^{q\alpha} e^\beta_q \sigma^\pm_{\alpha\beta}
$$

\noindent
The expression of the torsion ${\cal T}^{q\alpha}$ will not be needed in
the following and will not be reported here. It is only useful to recall
the Bianchi identity

$$
e^{q\alpha} e^{p\beta} e^{r\gamma} {\cal T}_{q\alpha,p\beta}\ ^{s\delta}
\sigma^\pm_{\delta\gamma} \delta_{rs} =0\eqno(2.2)
$$

The supermembrane target superspace, $\underline{\cal M}(11|32)$,
is described by the local supercoordinates:

$$
\hat Z^{\underline{\hat M}}\equiv
(X^{\underline{\hat m}},\theta^{\underline{\hat\mu}});\quad
{\underline{\hat m}}=0,1...10;\quad
\underline{\hat\mu}=1,...32.
$$

\noindent
We shall write $X^{10}=Y$ and we shall identify $Y$ with the
world volume coordinate $\rho$. Moreover a
Majorana spinor in D=11 decomposes in two Weyl--Majorana spinors in D=10
with opposite chirality. Then the string coordinates are

$$
Z^{\underline M}\equiv (X^{\underline m},\theta^{\underline\mu},
\bar\theta_{\underline\mu})\quad {\underline m}=0,1,...9;\quad
{\underline\mu}=1,...16
$$

\noindent
and $\hat Z^{\underline M}\equiv(Z^{\underline M}, Y)$. We assume that all
the background superfields do not depend on $Y$.

In D=11 the reonomic parametrizations of the torsion
$\hat T^{\underline{\hat A}}$
and of the $B$--curvature $\hat H=d\hat B$
are given
by

$$
\hat T^{\underline{\hat a}}= \hat E^{\underline{\hat\alpha}}
\hat E^{\underline{\hat\beta}} \hat\Gamma_{\underline{\hat\beta}
\underline{\hat\alpha}}^{\underline{\hat a}}\eqno(2.3a)
$$

$$\eqalign{
\hat T^{\underline{\hat\alpha}} = & {\sqrt{2}\over 3}
\hat E^{\underline{\hat\beta}} \hat E^{\underline{\hat a}}
\Bigl\{ F_{\underline{\hat a}\underline{\hat b}_1\underline{\hat b}_2
\underline{\hat b}_3}(\hat\Gamma^{{\underline b}_1 {\underline b}_2
{\underline b}_3}
)_{\underline{\hat\beta}}\ ^{\underline{\hat\alpha}} - {1\over 8}
F_{\underline{b}_1...\underline{b}_4}
(\hat\Gamma_{\underline{\hat a}}^{\underline{\hat b}_1...\underline{\hat b}_4}
)_{\underline{\hat\beta}}\ ^{\underline{\hat\alpha}}\Bigr\} + \cr
& + \hat E^{\underline{\hat a}} \hat E^{\underline{\hat b}}
\ \rho_{\underline{\hat b}\underline{\hat a}}\
^{\underline{\hat\alpha}}\cr}\eqno(2.3b)
$$

$$
\hat H = \hat E^{\underline{\hat\alpha}}
\hat E^{\underline{\hat\beta}} E^{\underline{\hat b}}
E^{\underline{\hat a}}
\Bigl(\hat\Gamma_{\underline{\hat a}\underline{\hat b}}\Bigr)_{
\underline{\hat\alpha}\underline{\hat\beta}} +
\hat E^{\underline{\hat a}}\hat E^{\underline{\hat b}}
\hat E^{\underline{\hat c}}\hat E^{\underline{\hat d}}
F_{\underline{\hat a}\underline{\hat b}\underline{\hat c}
\underline{\hat d}}\eqno(2.3c)
$$

The D=10 supervielbeins $E^A$, the Lorentz superconnection $\Omega_A\ ^B$
and the two--superform $B$ are obtained from the corresponding superforms
in D=11 by dimensional reduction. To this end we restrict all the target
superfields to be independent on $Y$ and we set:

$$
\hat E^{\underline a} = \bar e^\phi E^a\eqno(2.4a)
$$

$$
\hat E^{\hat{\underline\alpha}}= e^{-\phi/2}
\Bigl( E^{\hat{\underline\alpha}} + {1\over 2} E^{\underline b}
\Gamma_{\underline b}^{\hat{\underline\alpha}\hat{\underline\beta}}
\Delta_{\hat{\underline\beta}}\phi\Bigr) -
e^{\phi/2} \hat E^{\underline{10}} \Gamma_{\underline{10}}^{\hat{
\underline\alpha}\hat{\underline\beta}} \Delta_{\hat{\underline\beta}}
\phi\eqno(2.4b)
$$

$$
\hat E^{\underline{10}} = e^{2\phi} (E^{\underline {10}} + dY) =
e^{-\phi} (A+e^{\tilde\psi}\hat e^\perp)\eqno(2.4c)
$$

$$
\hat\Omega_{\underline a}\ ^{\underline b} =
\Omega_{\underline a}\ ^{\underline b} +
X_{\underline a}\ ^{\underline b}\eqno(2.4d)
$$

$$
\hat\Omega_{\underline{10}}\ ^{\underline b}
= X_{\underline{10}}\ ^{\underline b}
\eqno(2.4e)
$$

$$
B = dZ^{\underline M} dZ^{\underline N} B_{ {\underline N}
{\underline M} {\underline{10}}}\eqno(2.4f)
$$

\noindent
where

$$
E^{\underline a} = dZ^{\underline M} E^{\underline a}_{\underline M}
(Z);\quad
E^{\underline{10}}=dZ^{\underline M} E^{\underline{10}}_{\underline M}
(Z)
$$

$$
E^{\hat{\underline\alpha}} = dZ^{\underline M}
E_{\underline M}^{\hat{\underline\alpha}}(Z) =
(E^{\underline\alpha}, \bar E_{\underline\alpha})
$$

$$
A=e^{3\phi}
(E^{\underline{10}} - h)
$$

$$
\tilde\psi = \psi + 3\phi
$$

\noindent
$X^{\underline b}_{\underline{10}}$ and $X^{{\underline a}
{\underline b}} = - X^{{\underline b}{\underline a}}$ are Lorentz
covariant one--superforms, and $\phi(Z)$ is the dilaton.

One can see that a suitable choice of $X_{\underline{10}}\ ^{
\underline b}$ and $X_{\underline a}\ ^{\underline b}$ and the
identification of
$F_{{\hat{\underline a}}_1....{\hat{\underline a}}_4}$ and
$\rho^{\hat{\underline\alpha}}_{ {\underline{\hat a}} {\underline{\hat b}}}$ in
eq. (2.3b) with higher components of the superfield $\phi(Z)$ allow
to bring to zero the flat components of the torsion and Lorentz
curvature along $\hat E^{\underline {10}}$. Of course the dimensional
reduction is specified modulo a redefinition of the D=10 supervielbeins
and Lorentz connection and the choice in eqs. (2.2)--(2.4) is taken
to recover the standard
D=10 constraints$^{[18]}$. They are:

$$
T_{{\underline\alpha}{\underline\beta}}^{\underline a} =
2\Gamma_{{\underline\alpha}{\underline\beta}}^{\underline a};\quad
T^{{\underline a}{\alpha}{\beta}}=2
\Gamma^{{\underline a}{\underline\alpha}{\beta}}\eqno(2.5a)
$$

$$
T_{ {\underline\beta}{\underline\gamma}}\ ^{\alpha} = -  3
\Bigl( \delta_{\underline\beta}^{\underline\alpha}
\Delta_{\underline\gamma}\phi - {1\over 2}
\Gamma_{{\underline\beta}{\underline\gamma}}^{\underline b}
\Gamma_{\underline b}^{{\underline\alpha}{\underline\delta}}
\Delta_{\underline\delta}\phi\Bigr),\eqno(2.5b)
$$

$$
T^{ {\underline\beta}{\underline\gamma}}\ _{\underline\alpha} = - 3
\Bigr(\delta^{\underline\beta}_{\underline\alpha} \Delta^{\underline\gamma}
\phi - {1\over 2} \Gamma_{\underline b}^{{\underline\beta\gamma}}
\Gamma_{{\underline\alpha}{\underline\delta}}^{\underline b}
\bar\Delta^{\underline\delta} \phi\Bigr)\eqno(2.5c)
$$

$$
H_{{\underline a}{\underline\alpha}{\underline\beta}} =
\Gamma_{{\underline a}{\underline\alpha}{\underline\beta}} ; \quad
H_{\underline a}^{{\underline\alpha}{\underline\beta}} = -
\Gamma_{\underline a}^{{\underline\alpha}{\underline\beta}}\eqno(2.5d)
$$

$$
T_{{\underline b}{\underline c}}^{\underline a}=0\eqno(2.5e)
$$

\noindent
and all the other torsion and curvature components in the sectors of
dimensions 0, 1/2, vanish. One should notice that the torsion
components in eqs. (2.5 a,b,c) fulfil the Bianchi identity:

$$
T^{\underline\delta}_{({\underline\beta}{\underline\gamma}}
\Gamma^{\underline a}_{{\underline\alpha}){\underline\delta}}
= 0 \eqno(2.6)
$$

\noindent
Here and in the following, indices between round brackets (square
brackets) are symmetrized (antisymmetrized).

The pullback of $\hat E^{\hat{\underline A}}$
and $E^{\underline A}$ are respectively

$$
\hat E^{\underline{\hat A}} =
\hat e^+ \hat E_+^{\underline{\hat A}} +
\hat e^-\hat E_-^{\underline{\hat A}} +
\hat e^\perp  \hat E_\perp^{\underline{\hat A}} + \hat e^{q\alpha}
\hat E_{q\alpha}^{\underline{\hat A}}
$$

$$
E^{\underline A} = e^+ E_+^{\underline A} + e^-
E_-^{\underline A} +
e^{q\alpha}
E^{\underline A}_{q\alpha}
$$

\noindent
Then from eqs. (2.1) and (2.4 a,b,c) one has

$$
\hat E_A^{\underline a} = e^{-\phi} E^{\underline a}_A
$$

$$
\hat E^{\underline{10}}_A = A_A;\quad
\hat E^{\underline{10}}_\perp = e^{-\phi} e^{\tilde\psi}
$$

\noindent
The twistor constraint for the supermembrane is

$$
\hat E_{q\alpha}^{\hat{\underline a}}=0\eqno(2.7)
$$

\noindent
and the spinor--like derivative of this condition gives

$$
\Bigr( \hat E_{q\alpha} \hat\Gamma^{\underline{\hat a}}
\hat E_{p\beta}\Bigr) = \delta_{pq}
\Bigl[ \sigma^+_{\alpha\beta} \hat E_+^{\underline{\hat a}} +
\sigma^-_{\alpha\beta} \hat E_-^{\underline{\hat a}} +
\sigma^\perp_{\alpha\beta} \hat E_\perp^{\underline{\hat a}}\Bigr]
\eqno(2.8)
$$

\noindent
{}From eqs. (2.1) and (2.7) one gets

$$
\hat E^{\underline a}_\perp =  0
$$

\noindent
Moreover eq. (2.7) gives for the superstring the expected twistor constraint:

$$
E_{q\alpha}^{\underline a} = 0\eqno(2.9)
$$

\noindent
and the spinor--like derivative of eq. (2.9) (i.e. eq. (2.8)
restricted to D=10)  yields

$$
\delta_{qp} E_+^{\underline a} =
(E_{q1}\Gamma^{\underline a} E_{p1})+(\bar E_{q1}\Gamma^{\underline a}
\bar E_{p1})\eqno(2.10a)
$$

$$
\delta_{qp} E_-^{\underline a} = (E_{q2}\Gamma^{\underline a}
E_{p2})+(\bar E_{q2}\Gamma^a\bar E_{p2})\eqno(2.10b)
$$

$$
0= (E_{q1}
\Gamma^{\underline a} E_{p2})+(\bar E_{q1}\Gamma^{\underline a}
\bar E_{p2})\eqno(2.10c)
$$

\noindent
However eq. (2.8), taken for $\hat{\underline a}={\underline{10}}$,
contains the further constraints

$$
(\bar E_{(q1} E_{p1)})=\delta_{qp} A_+\eqno(2.11a)
$$

$$
(\bar E_{(q2} E_{p2)}) = \delta_{qp} A_-\eqno(2.11b)
$$

$$
(\bar E_{q1} E_{p2})+(\bar E_{p2} E_{q1})=\delta_{qp}
e^{\tilde\psi} \eqno(2.11c)
$$

\noindent
In conclusion as twistor constraints for the superstring we shall impose
both eq. (2.9) and eqs. (2.11). They will be obtained through lagrangian
multipliers by means of the action term:

$$
I^{(C)} = \int_{\cal M}\tilde e P^{q\alpha}_{\underline a}
E^{\underline a}_{q\alpha} + \int_{\cal M}\tilde e
Q^{\{q\alpha,p\beta\}} (\bar E_{q\alpha} E_{p\beta})\eqno(2.12)
$$

\noindent
where $Q^{\{q\alpha,p\beta\}}$ is symmetric in $(q\alpha),(p\beta)$
and traceless with respect to $q,p$, and $\tilde e$ is the
superdeterminant of $e_M^A$.

\vskip 0.5truecm

\noindent
{\bf 2 -- RELEVANT IDENTITIES AND WEYL TRIVIALITY}

\vskip 0.5truecm

As shown in [12], in the case of the supermembrane the twistor constraint
implies the remarkable identities

$$
V_{(qp)[\hat a \hat b]}^{\hat{\underline a}} =
\delta_{qp} V_{[\hat a \hat b]}^{\hat{\underline a}}\eqno(3.1)
$$

$$
V_{(qp)(\hat a \hat b)}^{\hat{\underline a}}=0\eqno(3.2)
$$

\noindent
and

$$
V_{(qp)\hat a \hat b}^{\hat{\underline a}}
\hat E_{\hat{\underline a}\hat c} = \delta_{qp}
\epsilon_{\hat a\hat b\hat c}\sqrt{det\ \hat g}\eqno(3.3)
$$

\noindent
where

$$
V_{(qp)\hat a \hat b}^{\hat{\underline a}} = {1\over 4}
\sigma_{\hat a}^{\alpha\beta} \sigma^{\gamma\delta}_{\hat b}
\Bigl( \hat E_{\alpha q}\Gamma^{\hat{\underline a}
{\hat{\underline b}}} \hat E_{\beta p}\Bigr)
\Bigl( \hat E_{\gamma r}\Gamma_{\hat b} \hat E_\delta^r\Bigr)
\eqno(3.4)
$$

\noindent
Here $q,p=1,...,n$; $3\leq n \leq 8$ and $\hat a,\hat b = \pm, \perp$.

 Moreover

$$
\hat g_{\hat a \hat b} = \hat E_{\hat a}^{\hat{\underline a}}
\eta_{{\hat{\underline a}}
{\hat{\underline b}}} \hat E^{\hat{\underline b}}_{\hat b}\eqno(3.5)
$$

\noindent
is the metric induced by the target vielbeins $\hat E^{\hat{\underline a}}$
in the tangent space of the world volume. We shall assume that this metric
is non degenerate.

Let us consider the projectors

$$
\hat Q^\pm={1\over 2} (1\pm {\hat{\bar\Gamma}})
$$

\noindent
where

$$
{\hat{\bar\Gamma}} =
{\epsilon^{\hat a\hat b\hat c} \hat E_{\hat a}^{\hat{\underline a}}
\hat E^{\hat{\underline b}}_{\hat b}
\hat E^{\hat{\underline c}}_{\hat c}
\Gamma_{ {\hat{\underline a}} {\hat{\underline b}}{\hat{\underline c}} }
\over 6\sqrt{det\ \hat g}}\eqno(3.6)
$$

\noindent
and let us write

$$
\hat v^{(\pm)} = \hat Q^\pm \hat v
$$

\noindent
where $\hat v$ is a D=11 spinor. It is also convenient to split the
$\Gamma$--matrices as

$$
\Gamma^{\hat{\underline a}} =
\Gamma_=^{\hat{\underline a}} + \Gamma_\perp^{\hat{\underline a}}
$$

\noindent
where $\Gamma_=^{\hat{\underline a}}$ lives in the three dimensional
subspace spanned by $E_a^{\hat{\underline a}}$ and
$\Gamma_\perp^{\hat{\underline a}}$ lives in the eight dimensional
orthogonal subspace. Notice that

$$
(\hat v^{(\pm)} \Gamma_\perp^{\hat{\underline a}} \hat v^{(\pm)})
=0= (\hat v^{(\pm)} \Gamma_=^{\hat{\underline a}} \hat v^{(\mp)})
$$

\noindent
Eq. (3.3) is equivalent to the following condition

$$
(\hat E_{(q\alpha}^{(-)}\Gamma^{\hat{\underline a}}
\hat E_{p\beta)})=0\eqno(3.7)
$$

\noindent
that is

$$
\sigma_a^{\alpha\beta}(\hat E_{(q\alpha}^{(-)}
\Gamma_=^{\hat{\underline a}}\hat E_{p\beta)}^{(-)})=0\eqno(3.8)
$$

$$
\sigma_a^{\alpha\beta}(\hat E_{(q\alpha}^{(-)}
\Gamma_\perp^{\hat{\underline a}}\hat E_{p\beta)}^{(+)})=0\eqno(3.9)
$$

After dimensional reduction from D=11 to D=10, eq. (3.3) yields
the identities

$$\eqalign{
\Bigl[ (\bar E_{q1}\Gamma^{\underline a}\bar E_{p1})&-(E_{q1}
\Gamma^{\underline a} E_{p1})\Bigr] E_{-\underline a} = \cr
=\Bigl[ (E_{q2} \Gamma^{\underline a} E_{p2})&-(\bar E_{q2}
\Gamma^a\bar E_{p2})\Bigr] E_{+{\underline a}}=
\delta_{qp}(- det\ g)^{1/2}.\cr}\eqno(3.10)
$$

$$
{1\over 2} \Bigl[
(E_{q1} \Gamma^{\underline a} E_{p2}) +
(E_{q2}\Gamma^{\underline a} E_{p1}) - (\bar E_{q1} \Gamma^{\underline a}
\bar E_{p2}) - (\bar E_{q2}\Gamma^{\underline a} \bar E_{p1})\Bigr]
(A_+ E^{\underline a}_- - A_-E^{\underline a}_+)=
$$

$$
= {1\over 2} \Bigl[ (E_{q1}\Gamma^{\underline{ab}}\bar E_{p2})+
(E_{q2} \Gamma^{\underline{ab}}\bar E_{p1})\Bigr]
E_{+{\underline a}}E_{-{\underline b}}
= e^{\tilde\psi} \delta_{qp} (- det\ g)^{1/2}\eqno(3.11)
$$

\noindent
and eq. (3.7) gives

$$
(E_{(q\alpha}^{(-)} \Gamma^{\underline a} E_{p\beta)}) +
(\bar E_{(q\alpha}^{(-)} \Gamma^{\underline a}
\bar E_{p\beta)})=0\eqno(3.12)
$$

\noindent
Here

$$
g_{ab} = E^{\underline a}_a \eta_{\underline{ab}}
E^{\underline b}_b\eqno(3.13)
$$

\noindent
and

$$
E^{(\pm)}_{q\alpha} = Q^\pm E_{q\alpha} ;\quad
\bar E^{(\pm)}_{q\alpha}=Q^\pm \bar E_{q\alpha}\eqno(3.14)
$$

\noindent
with

$$
Q^\pm = {1\over 2}
{ (1 \pm E_+^{\underline a} E_-^{\underline b}
\Gamma_{\underline{a}\underline{b}})\over (det\ g)^{1/2} }\eqno(3.15)
$$

\noindent
Moreover eqs. (3.2), (3.4), taken for $\hat a=\hat b=\perp$
and $q=p$ together with eq. (2.11c) imply

$$
\Bigl[ (E_{q1}\Gamma^{\underline a} E_{q2})-
(\bar E_{q1}\Gamma^{\underline a} \bar E_{q2})\Bigr]=0\eqno(3.16)
$$

\noindent
so that, by taking into account eq. (2.10c), one has

$$
(E_{q1} \Gamma^{\underline a} E_{q2}) = 0 =
(\bar E_{q1} \Gamma^{\underline a} \bar E_{q2})\eqno(3.17)
$$

Of course these identities can be derived directly from the twistor
constraints in D=10.

Indeed let us consider the vector

$$\eqalign{
V_q^{\underline a} & =(E_{q1} \Gamma^{\underline a{\underline b}}
\bar E_{q1}) E_{-{\underline b}} + (E_{q2} \Gamma^{{\underline a}
{\underline b}} \bar E_{q2}) E_{+{\underline b}} -
\Bigl[ (E_{2q}\Gamma^{\underline a} E_{2q})-
(\bar E_{2q}\Gamma^{\underline a}\bar E_{2q})\Bigr] A_+ -\cr
& - [(E_{q1} \Gamma^{\underline a} E_{q1})-(\bar E_{q1}
\Gamma^{\underline a}\bar E_{q1})] A_-\cr}\eqno(3.18)
$$

Then a straightforward calculation, which makes use of the
constraints (2.10), (2.11) and of the $\Gamma$--matrix cyclic
identity, allows to rewrite $V_q^{\underline a}$ as

$$
V^{\underline a}_q =-4\Bigl[(E_{q1}\bar E_{p2})
+(\bar E_{q1} E_{p2})\Bigr]
(\bar E_{q1} \Gamma^{\underline a} \bar E_{p2}) + W_q^{\underline a}
- W_p^{\underline a}\eqno(3.19)
$$

\noindent
where

$$
W_q^{\underline a} = (E_{q2}\Gamma^{{\underline a}{\underline b}}
\bar E_{q2}) E_{+\underline b} - [(E_{2q}\Gamma^{\underline a} E_{2q})-
(\bar E_{2q} \Gamma^{\underline a}\bar E_{2q})]A_+
$$

\noindent
{}From eq. (3.19), if $q\not =p$ one has

$$
V_q^{\underline a} = W_q^{\underline a} -
W_p^{\underline a}
$$

\noindent
so that, for $q\not = p \not = r$,

$$
V_p^{\underline a} + V_q^{\underline a}
=0 = V_q^{\underline a} + V_r^{\underline a}
$$

\noindent
and therefore $V_q^{\underline a}=0$. Then, considering eq. (3.19)
for $q=p$, one obtains eq. (3.17). Notice that in this
derivation $n$ is required to be $\geq 3$.

An interesting consequence
of eq. (3.17) is that the vectors

$$
L^{\underline a}_{q\pm} =
\sigma^{\alpha\beta}_\pm
(E_{q\alpha}
\Gamma^{\underline a} E_{q\beta})\eqno(3.20)
$$

\noindent
are proportional. Indeed they are light--like and moreover, from
the cyclic identity and eq. (3.12),

$$
L^{\underline a}_{q+} L_{q-,{\underline a}} =0
$$

\noindent
In the same way, also $\bar L_{q+}^{\underline a}=(\bar E_{q1}
\Gamma^{\underline a} \bar E_{q1})$ and $\bar L_{q-}^{\underline a}=
(\bar E_{q2}\Gamma^{\underline a} \bar E_{q2})$ are proportional.

Moreover the proportionality coefficients as well as the scalar products
$L^{\underline a}_{q\pm} \bar L_{q\pm{\underline a}}$ are independent from $q$.
This can be seen by expressing the metric components $g_{ab}$ in eq.
(3.5) in terms of $L^{\underline a}_{q\pm}$, $\bar L^{\underline a}_{
q\pm}$. Indeed, from eqs. (2.10), $L^{\underline a}_{q\pm}$ and
$\bar L^{\underline a}_{\pm}$ are $q$--independent.

\noindent
Then we can write

$$
L^{\underline a}_{q\pm} = \alpha_\pm L^{\underline a}\qquad
\bar L^{\underline a}_{q\pm} = \beta_\pm \bar L^{\underline a}
\eqno(3.21)
$$

\noindent
and we can set, without restriction, $| det\ A| =1$, where $A$ is the
invertible matrix $A={\alpha_+\ \beta_+\choose \alpha_-\ \beta_-}$.

\noindent
It is also convenient to write $\det\ g$ in terms of these vectors.
One gets

$$
- det\ g =
(L^{\underline a}\bar L_{\underline a})^2\eqno(3.22)
$$

At this point, it is immediate to verify eqs. (3.10) and (3.11) and
then to deduce eq. (3.12).

Restricting ourselves to the case $n=8$, let us notice that the
16$\times$16 matrix $E^{(+)\hat{\underline\alpha}}_{q\alpha}\equiv
(E^{(+){\underline\alpha}}_{q\alpha},
\bar E^{(+)}_{q\alpha{\underline\alpha}})$,
is invertible.

Indeed its inverse is $F^{(+)q\alpha}_{\hat{\underline\alpha}} =
(F^{(+)q\alpha}_{\underline\alpha},
\bar F^{(+)q\alpha{\underline\alpha}})$
where

$$\eqalign{
F^{(+)q\alpha}_{\underline\alpha} & = {1\over 16}
\sigma_a^{\alpha\beta} g^{ab} E_b^{\underline a} (\Gamma_{\underline a}
E^q_\beta)_{\underline\alpha}\cr
\bar F^{(+)q\alpha{\underline\alpha}} & = {1\over 16}
(\sigma_a)^{\alpha\beta} g^{ab} E_b^{\underline a}
(\Gamma_{\underline a} \bar E^q_\beta)^{\underline\alpha}\cr}
$$

\noindent
so that the projector  $Q^{(+)}$ can be rewritten as

$$
Q^{(+)\hat{\underline\beta}}_{\hat{\underline\alpha}}=
F^{(+)q\alpha}_{\hat{\underline\alpha}}
E^{(+)\hat{\underline\beta}}_{q\alpha}\eqno(3.23)
$$

Now we are ready to prove Weyl triviality for the 2--superform
$B$.

Weyl triviality asserts that it is possible to modify $B$
by adding to it a gauge and Lorentz invariant 2--superform $K$,

$$
\tilde B = B+K\eqno(3.24)
$$

\noindent
in such a way that the differential of $\tilde B$ restricted to the
superworldsheet ${\cal M}$ vanishes. In our case

$$
K={1\over 4 n} e^+ e^-
\Bigl( E_-^{\underline A} \sigma_+^{\alpha\beta} - E_+^{\underline A}
\sigma_-^{\alpha\beta}\Bigr)
E^{\underline B}_{q\alpha} E^{q{\underline C}}_\beta
H_{{\underline C}{\underline B}{\underline A}} \eqno(3.25)
$$

Using the twistor constraints and the SUGRA constraints for $H$, as well
as eqs.(3.10), (3.22), $K$ can be rewritten as

$$
K = {1\over 2}  e^+ e^- (- det\ g)^{1/2} =
{1\over 2} d\xi^1\ d\xi^2 (- det\ G)^{1/2}\eqno(3.26)
$$

\noindent
where

$$
G_{mn} = E^{\underline a}_m \eta_{\underline{ab}}
E^{\underline b}_n\eqno(3.27)
$$

By taking the differential of eq. (3.25) (and using eq. (3.10)) one has

$$
dK  = {1\over 2}\Bigl[(\sigma^+)_{\alpha\beta} e^- e^\alpha_q
e^{q\beta}
-(\sigma^-)_{\alpha
\beta} e^+ e^\alpha_q e^{q\beta}
\Bigr] (-det\ g)^{1/2} +
$$
$$
+ e^+ e^- e^{q\alpha}
\Bigl[ (E^{\underline a}_+ E^{\underline\alpha}_- - E_-^{\underline a}
E^{\underline\alpha}_+) \Gamma_{{\underline a}{\underline\alpha}
{\underline\beta}} E^{\underline\beta}_{q\alpha} -
(E_+^{\underline a} \bar E_{-{\underline\alpha}} -
(E^{\underline a}_- \bar E_{+{\underline\alpha}})
\Gamma_{\underline a}^{{\underline\alpha}{\underline\beta}}
\bar E_{q\alpha{\underline\beta}}\Bigr]\eqno(3.28)
$$

\noindent
To get eq. (3.28) one should remark that the contributions to $dK$
involving the components ${\cal T}^{q\alpha}_{p\beta,r\gamma}$ and
$T^{\underline\alpha}_{{\underline\beta}{\underline\gamma}},
T^{{\underline\beta}{\underline\gamma}}_{\underline\alpha}$ of the
worldsheet   and target space torsions, vanish. Indeed, if for
instance one considers the term

$$
C= E_-^{\underline a}\sigma_+^{\alpha\beta} \Delta_{r\gamma}
(E_{q\alpha} \Gamma^{\underline a} E^q_\beta)
$$

\noindent
which arise in the calculation of $dK$ one gets

$$
C=E_-^{\underline a} (E_+\Gamma^a E_{r\gamma})
+ E_-^{\underline\alpha} \sigma_+^{\alpha\beta}
\Bigl[ {\cal T}^\delta_{(r\gamma,q\alpha,p}
(\sigma^+)_{\beta)\delta} L_+^{\underline a} +
E^{\underline\gamma}_{r\gamma}
E^{\underline\alpha}_{q\alpha}
E^{\underline\beta}_{p\beta}
T^{\underline\delta}_{({\underline\gamma}{\underline\beta}}
\Gamma^{\underline\alpha}_{\underline\alpha)\delta} \Bigr]
\delta^{pq}
$$

\noindent
and the second term vanish due to
the Bianchi identities eqs. (2.2),
(2.6).

\noindent
On the other hand the pull back of $dB$ is

$$
dB\mid_{\cal M} = \Bigl\{ {1\over 2} e^{q\alpha} e^{p\beta}
(e^+ E_+^{\underline a} + e^- E_-^{\underline a})
\Bigl[ (E_{q\alpha}\Gamma_{\underline a} E_{p\beta}) -
(\bar E_{q\alpha} \Gamma_{\underline a}\bar E_{p\beta})\Bigr]-
$$
$$
- e^{q\alpha} e^+ e^- \Bigl[ E_+^{\underline a}
\Bigl( (E_-\Gamma_{\underline a} E_{q\alpha})-(\bar E_-
\Gamma_{\underline a} \bar E_{q\alpha})\Bigr) - E^{\underline a}_-
\Bigl( (E_+\Gamma_{\underline a} E_{q\alpha}) - (\bar E_+
\Gamma_{\underline a}E_{q\alpha})\Bigr)\Bigr] \Bigr\}
\Big\vert_{\cal M}\eqno(3.29)
$$

However, from eqs. (2.10),
(3.21), (3.22) and the cyclic identity, one has

$$
E^{\underline a}_\pm \Bigl[ (E_{q\alpha} \Gamma_{\underline a}
E_{p\beta})-(\bar E_{q\alpha} \Gamma_{\underline a} \bar E_{p\beta})
\Bigr] = \pm (- det\ g)^{1/2} \delta_{qp}
(\sigma_\pm)_{\alpha\beta}
$$

\noindent
so that the r.h.s. of eqs. (3.28) and (3.29) are equal and opposite
and Weyl triviality
is proved.

\vskip 0.5truecm

\noindent
{\bf 4 -- ACTION AND FIELD EQUATIONS}

\vskip 0.5truecm

\noindent
It follows from Weyl triviality that the action

$$
I^{(B)}=\alpha \int_{{\cal M}_0} \tilde B\eqno(4.1)
$$

\noindent
is invariant under $n$--extended worldsheet supersymmetry. Here
${\cal M}_0$ is the slide of ${\cal M}$ at $\eta^{qn}=0$, the constant
$\alpha$ is the string tension and

$$
\tilde B=B + {1\over 4n} e^+ e^- \epsilon^{ab}
\sigma^{\alpha\beta}_b
E^{\underline A}_{q\alpha}
E_\beta^{q{\underline B}} E^C_a H_{\underline{CBA}}\eqno(4.2)
$$

Indeed, if $\delta_\epsilon$ denotes the variation under the infinitesimal
local supersymmetry transformation $\zeta^M\rightarrow\zeta^M +
\epsilon^{q\alpha}(\zeta) e^M_{q\alpha}$ and $i_\epsilon$ denotes the
contraction of a (super)form with the vector $\epsilon^{q\alpha}
e^M_{q\alpha}$, one has

$$
\delta_\epsilon\tilde B = di_\epsilon \tilde B + i_\epsilon d\tilde B = d
i_\epsilon\tilde B
$$

\noindent
so that $\delta_\epsilon I^{(B)}=0$.

The action $I^{(B)}$ must be added to the action $I^{(C)}$ in eq.
(2.12). Alternatively one can add to $I^{(C)}$ the action $^{[9]}$

$$
I'^{(B)} = \int_{\cal M} P^{MN} (\tilde B_{NM}-
\partial_N Q_M)\eqno(4.3)
$$

\noindent
where $P^{MN}$ are new, Grassmann antisymmetric, lagrangian multipliers.
The local invariance, that follows from Weyl triviality,

$$
\delta P^{MN}=\partial_L\Lambda^{LMN},
$$

\noindent
$\Lambda^{LMN}$ being a Grassmann antisymmetric superfield, allows to
gauge to zero all the components of $P^{MN}$ excepting the highest one
$^{[9][19]}$.

$$
P^{m\ell} =\alpha(\eta^2)^n \epsilon^{m\ell}
$$

\noindent
where

$$
\eta^2 = {1\over 2n} (\eta^{q\mu} \eta^\nu_q \epsilon_{\mu\nu})
$$

\noindent
so that eq. (4.1) is recovered.

In order to implement the worldsheet supervielbeins and torsion
constraints one can add to the action $I^{(B)} + I^{(C)}$ the
further term

$$\eqalign{
I^{(T)} = & \int_{\cal M}\tilde e \Bigl[ K_a^{q\alpha,p\beta}
\Bigl( {\cal T}^a_{q\alpha, p\beta} - \sigma^a_{\alpha\beta}
\delta_{qp}\Bigr) + K^{p\beta}_{q\alpha}
\Bigl( e_{p\beta} e^{q\alpha} - \delta^q_p \delta^\alpha_\beta\Bigr)+\cr
& + K^b_a (e_b e^a-\delta^a_b) + K^b_{q\alpha} (e_b e^{q\alpha})+
K_a^{p\beta} (e_{p\beta} e^a)\Bigr]\cr}\eqno(4.4)
$$

Finally let us recall that by a shift of the lagrangian multipliers
$P^{\alpha q}_{\underline a}$, $I^{(B)}$ can be rewritten as

$$
I^{(B)} = \int d^2\xi \Bigl(\epsilon^{mn} B_{mn}
 + (- det\ G)^{1/2}\Bigr)
\eqno(4.5)
$$

\noindent
and reduces to a form of the standard action of the G.S., type IIA,
superstring, $\sigma$--model.

In conclusion the twistor--like formulation of the type II A,
superstring $\sigma$--model is described by the following action

$$
I=\int d^2\xi [\epsilon^{mn}B_{mn} + (- det\ G)^{1/2}] +
\int_{\cal M}\tilde e \Bigl[P_{\underline a}^{q\alpha}
E_{q\alpha}^{\underline a} + Q^{ \{q\alpha,p\beta\}}
\Bigl((E_{(q\alpha} \bar E_{p\beta})\Bigr)\Bigr] + I^{(T)}\eqno(4.6)
$$

\noindent
(Here $P^{q\alpha}$ and $Q^{ \{q\alpha,p\beta\}}$ denote the shifted
lagrangian multipliers).

The relevant field equations are

$$
\Delta_{q\alpha} P_{\underline a}^{q\alpha}+(\eta^2)^n
{\cal L}_{\underline a}=0\eqno(4.7)
$$

$$
P^{q\alpha}_{\underline a} (\Gamma^{\underline a}
E_{q\alpha})_{\underline\alpha} - \Delta_{q\alpha}
(Q^{\{q\alpha,p\beta\}}\bar E_{p\beta{\underline\alpha}})
+ (\eta^2)^n {\cal L}_{\underline\alpha} =0\eqno(4.8)
$$

$$
P^{q\alpha}_{\underline a} (\Gamma^{\underline a}
\bar E_{q\alpha})^{\underline\alpha} - \Delta_{q\alpha}
(Q^{\{q\alpha,p\beta\}}E_{p\beta}^{\underline\alpha})
+ (\eta^2)^n {\cal L}^{\underline\alpha} =0\eqno(4.9)
$$

$$
P^{q\alpha}_{\underline a} E_b^{\underline a} +
Q^{ \{q\alpha,p\beta\}} [(E_b\bar E_{p\beta})+(\bar E_b
E_{p\beta})] - \Delta_{p\beta}
K_b^{\{q\alpha, p\beta\}} =0\eqno(4.10)
$$

$$
K^{p\beta}_{q\alpha}=0=K^b_a;\qquad K^b_{q\alpha}=0\eqno(4.11)
$$

$$
K_a^{q\alpha}=\Delta_{p\beta} K_a^{\{q\alpha, p\beta\}}\eqno(4.12)
$$

\noindent
where

$$
{\cal L}^{\underline A} =
E^{\underline A}_{\underline M}
{\delta I^{(B)}\over\delta Z^{\underline M}}=0\eqno(4.13)
$$

\noindent
are the standard superstring field equations. Moreover the only
non vanishing components of $K_a^{\{q\alpha,p\beta\}}$ are
$K_+^{\{q1,p1\}}$ and
$K_-^{ \{q2,p2\}}$ symmetric and traceless in $q,p$.

The action $I$ is invariant not only under diffeomorphisms and
$n$--extended local supersymmetry but also under the generalized
superWeyl transformations

$$
e'^\pm = e^{\Lambda(\zeta)} e^\pm
$$

$$
e'^{q\alpha} = e^{\Lambda(\zeta)/2} e^{q\alpha} + e^\pm
\Lambda_\pm^{q\alpha}(\zeta)
$$

\noindent
supplemented with corresponding transformations of $D_A$ and a
rescaling of the lagrangian multipliers. In addition I is invariant
under the following local
transformations of the lagrangian multipliers

$$
\delta^{(1)} P^{q\alpha\ {\underline a}} =
\Delta_{p\beta} \Lambda_{\underline b}^{ \{q\alpha,p\beta,r\gamma,
s\delta\}} \Bigl(\bar E_{r\gamma} \Gamma^{ {\underline b}{\underline a}}
E_{s\delta}\Bigr)+ \dots\eqno(4.14a)
$$

$$
\delta^{(1)} Q^{\{q\alpha,p\beta\}} =
\Lambda_{\underline b}^{\{q\alpha,p\beta,r\gamma,s\delta\}}
\Bigl[ (E_{r\gamma} \Gamma^{\underline b} E_{s\delta})-(\bar E_{r\gamma}
\Gamma^{\underline b} \bar E_{s\delta})\Bigr]+ ....\eqno(4.14b)
$$

$$
\delta^{(2)} Q^{\{q\alpha,p\beta\}} =
\Delta_{\gamma r} \Lambda ^{\{q \alpha, p\beta, r\gamma\}}+....\eqno(4.14c)
$$

$$
\delta^{(3)} P^{q\alpha}_{\underline a} = \Delta_{p\beta}
\Bigl[ \Delta_{r\gamma} \Lambda^{r\gamma \{q,p\}}_{\{a,b\}}
(\sigma^b)^{\alpha\beta}\ g^{af} E_{f{\underline a}}\Bigr]+\cdots
\eqno(4.15)
$$

$$
\delta^{(4)} K_a^{q\alpha,p\beta} = \Delta_{r\gamma}
\Lambda_{\{a,b,c\}\delta}^{\{r,q,p\}}
(\sigma^b)^{\alpha\beta} (\sigma^c)^{\gamma\delta}+....\eqno(4.16)
$$

\noindent
Indices between curly brackets are symmetrized and traceless and
in particular $\Lambda_{\underline b}^{\{q\alpha,p\beta,r\gamma,s\delta\}}$
are superfields symmetric in $q\alpha, p\beta,r\gamma,s\delta$ and
traceless
in $q,p,r,s,$. The dots in eqs. (4.14a), (4.14c) and (4.15)
denote suitable terms proportional
to the gauge parameters $\Lambda^{\prime s}$ and/or
${\cal T}^{r\gamma}_{q\alpha,p\beta}$.
The invariance under (4.14) and (4.16) follows
immediately from the cyclic identity and the torsion Bianchi identity
respectively. That under (4.15) is less obvious. It encodes the fact that
the spinor like derivatives of the components of the constraints (2.10)
parallel to $E_{a{\underline a}}$ and traceless in $q,p$ and of the
constraints (2.11) traceless in $q,p$ are not new constraints but are
fulfilled automatically once the constraints (2.9), (2.10), (2.11) are
satisfied. This can be shown easy using eqs. (3.13) and (3.28).

Coming
back to the field equations (4.7)--(4.10), eq. (4.7) implies

$$
P_{\underline a}^{q\alpha} =
\Delta_{p\beta} \Bigl[ \bar Q_{\underline a}^{ \{q\alpha,p\beta\}}+
(\eta^2)^n \delta^{qp} \sigma^{\alpha\beta}_a p^a_{\underline a}\Bigr]
\eqno(4.17)
$$

Moreover from eqs, (4.8) and (4.9) one can see that the superfields
$\Delta_{p\beta} \bar Q_{\underline a}^{ \{q\alpha,p\beta\}}$ and
$Q^{ \{q\alpha,p\beta\}}$ have the
same structure of
the r.h.s. of eqs. (4.14) and (4.15) so that they can be gauged away.
In addition eqs. (4.8) and (4.5) require that the vectors
$p^\pm_{\underline a}$ in eq. (4.17) are proportional both to
$L^{\underline a}$ and to $\bar L^{\underline a}$ so
that they vanish.
Then
eq. (4.10) together with eq. (4.16), allows to eliminate
$K_a^{q\alpha, p\beta}$ as well. Now eqs. (4.7), (4.8), (4.9) reduce to
the classical field equations of the standard, Green--Schwarz,
type IIA superstring $\sigma$--model

$$
{\cal L}_{\underline a} =0
$$

$$
{\cal L}_{\underline \alpha} = 0 = {\cal L}^{\underline\alpha}
$$

\noindent
It is worth mentioning that the components of the supervielbeins and
$SO(1,1)\otimes SO(n)$ connection are not independent dynamical variables.
Indeed $\omega_A\ ^B$ can be expressed in terms of the supervielbeins
by means of conventional torsion constraints and the superdiffeomorphisms
and the other local invariances allow to gauge away almost all
supervielbeins components. Moreover the (derivatives of the) twistor
constraints allow to express the remaining components such as the gravitino
fields (as well as the higher components of the superfields $Z^M$) in
terms of the leading components of the ``higher" superfields and of the
twistor fields.

Let us conclude by noticing explicitly a fact already anticipated in the
introduction. The induced metric

$$
g_{ab}=E_a^{\underline a} \eta_{{\underline a}{\underline b}}
E_b^{\underline b}
$$

\noindent
is not diagonal and therefore the Virasoro constraints are not
fulfilled in the frame where the n--extended supersymmetry is manifest.
Nevertheless, it is easy to recover the frame where the Virasoro
constraints hold recalling eqs. (3.20)--(3.22).
Indeed eqs. (2.10 a,b) can be rewritten in the form

$$
{E^{\underline a}_+\choose E_-^{\underline a}}\qquad = A
{L^{\underline a}\choose \bar L^{\underline a}}
$$

\noindent
Then the required frame is

$$
{e'_+\choose e'_-} = A^{-1}
{e_+\choose e_-}
$$

\noindent
In this frame $E^{\prime{\underline a}}_+ = L^{\underline a}$
and $E^{\prime{\underline a}}_-=\bar L^{\underline a}$
so that

$$
E^{\prime{\underline a}}_+ E^\prime_{+{\underline a}} = 0 =
E^{\prime{\underline a}}_- E^\prime_{-{\underline a}}$$

\noindent
However in this frame the worldsheet supersymmetry is hidden.

\vskip 0.5truecm

\noindent
{\bf Acknowledgements}. We are grateful to D. Sorokin for helpful
comments and discussions and to E. Sezgin for useful comunications.

\vskip 0.5truecm

\noindent
{\bf Appendix}

\vskip 0.5truecm

In the target space the tangent Minkowski metrics $\hat\eta_{
\hat{\underline a}\hat{\underline b}}$ in D=11 and $\eta_{{\underline a}
{\underline b}}$ in D=10 are

$$
\hat\eta_{{\underline 0}{\underline  0}} = -
\hat\eta_{{\underline a}'{\underline a}'} = -
\hat\eta_{{\underline {10}}{\underline{10}}} = 1;\quad
\eta_{{\underline 0}{\underline 0}}=-
\eta_{ {\underline a}'{\underline a}'}=1
$$

\noindent
where

$$
\hat{\underline a} = ({\underline a},{\underline{10}})\ ;\
{\underline a} = ({\underline 0}, {\underline a}')\ {\rm and}\
{\underline a}' = 1,...9
$$

\noindent
The Dirac matrices in D=11 are 32$\times$32 matrices for which we shall
choose the representation

$$
\hat\gamma^{\underline 0} = {\bf 1}_{16}\otimes\tau_2;\quad
\hat\gamma^{{\underline a}'}=(-i)\gamma^{a'}\otimes\tau_1;\quad
\hat\gamma^{\underline{10}}= i{\bf 1}_{16}\otimes\tau_3;\eqno(A1)
$$

\noindent
where $\tau_k, k=1,2,3$ are the Pauli matrices and the nine real
16$\otimes$16 matrices $\gamma^{{\underline a}'}$ satisfy the Clifford
algebra.

The real and symmetric $\Gamma$--matrices are defined as

$$
( \hat\Gamma^{\hat{\underline
a}})_{\hat{\underline\alpha}\hat{\underline\beta}}
=i (\hat\gamma^{\hat{\underline a}} \hat C)_{ {\hat{\underline\alpha}}
{\hat {\underline\beta}}}
$$

\noindent
where $\hat C=-i{\bf 1}_{16}\otimes\tau_2$ is the charge conjugation
matrix. The $\Gamma$--matrices satisfy the cyclic identity

$$
(\hat\Gamma^{\hat{\underline a}})_{
(\hat{\underline\alpha}\hat{\underline\beta}}
(\hat\Gamma_{\hat{\underline a}{\hat{\underline b}}}
)_{ \hat{\underline\gamma}
 \hat{\underline\delta)}}=0
$$

\noindent
and in the rapresentation (A1) are

$$\eqalign{
\hat\Gamma^{\underline 0} & = {\bf 1}_{16}\otimes{\bf 1}_2 =
{{\bf 1}\ 0\choose 0\ {\bf 1}}\cr
\hat\Gamma^{{\underline a}'}& = \gamma^{\underline a'}\otimes \tau_3 =
{\gamma^{\underline a'}\ 0\choose 0\ -\gamma^{\underline a'}}\cr
\hat\Gamma^{\underline{10}} & ={\bf 1}_{16}\otimes\tau_1 =
{0\ {\bf 1}\choose {\bf 1}\ 0}\cr}\eqno(A2)
$$

\noindent
Similarly, we define

$$
(\hat\Gamma^{\hat{\underline a}_1...\hat{\underline a}_k}
)_{\hat{\underline\alpha}\hat{\underline\beta}}=(i)^k
(\hat\gamma^{[\hat{\underline a}_1}\cdots\hat\gamma^{\hat{
\underline a}_k]}\hat C)_{\hat{\underline\alpha}\hat{\underline\beta}}
$$

\noindent
In particular

$$
\hat\Gamma^{{\underline a}'{\underline b}'} =
\left(\matrix{
0 & \gamma^{ [{{\underline a}'}} \gamma^{ {{\underline b}'}]}\cr
-\gamma^{ {[{\underline a}'}}\gamma^{{\underline b}']} & 0\cr}\right);
\hat\Gamma^{{{\underline a}'}{\underline{10}}} =
\left(\matrix{
\gamma^{{\underline a}'} & 0\cr
0   &\gamma^{{\underline a}'}\cr}\right)
$$

$$
\hat\Gamma^{{{\underline a}'}{\underline 0}} =
\left(\matrix{
0 & \gamma^{{\underline a}'}\cr
\gamma^{{\underline a}'} & 0\cr}\right) ;
\hat\Gamma^{{\underline 0}{\underline{10}}} =
\left(\matrix{
{\bf 1} & 0\cr
0 & {\bf 1}\cr}\right)\eqno(A3)
$$

A Majorana spinor in D=11 reduces in D=10 to a couple of
Weyl--Majorana in D=10 with opposite chiralities:
$\hat v^{\hat{\underline a}}\equiv(v^{\underline\alpha},
\bar v_{\underline\alpha})$. Then from eqs. (A1)--(A3)

$$
(\hat\Gamma^{\underline a})_{{\hat{\underline\alpha}}
{\hat{\underline\beta}}} =
\left(\matrix{
(\Gamma^{\underline a})_{{\underline\alpha}{\underline\beta}} & 0\cr
0 & (\Gamma^{\underline a})^{{\underline\alpha}{\underline\beta}}
\cr}\right) ; \quad
(\hat\Gamma^{\underline{10}})_{{\hat{\underline\alpha}}
{\hat{\underline\beta}}} =
\left(\matrix{
0 & {\bf 1}_{\underline\alpha}\ ^{\underline\beta}\cr
{\bf 1}^{\underline\alpha}\ _{\underline\beta} & 0\cr}\right)
$$

$$
(\hat\Gamma^{{\underline a}{\underline b}})_{ \hat{\underline\alpha}
\hat{\underline\beta}} =
\left(\matrix{
0  & (\Gamma^{{\underline ab}})_{\underline\alpha}\ ^{\underline\beta}\cr
(\Gamma^{\underline{ab}})^{\underline\alpha}\ _{\underline\beta} & 0\cr}\right)
;\quad
(\hat\Gamma^{{\underline a}\underline{10}})_{{\hat{\underline\alpha}}
{\hat{\underline\beta}}} =
\left(\matrix{
(\Gamma^{\underline a})_{{\underline\alpha}{\underline\beta}} & 0\cr
0 & -(\Gamma^a)^{{\underline\alpha}{\underline\beta}}\cr}\right)
$$

\noindent
$(\Gamma^{\underline a})_{{\underline\alpha}{\underline\beta}}$ and
$(\Gamma^{\underline a})^{{\underline\alpha}{\underline\beta}}$ are
the real and symmetric $\Gamma$--matrices in D=10, acting on chiral
and antichiral spinors respectively. They satisfy the cyclic
identities

$$
(\Gamma^{\underline a})_{{\underline\alpha}({\underline\beta}}
(\Gamma_{\underline a})_{{\underline\gamma}{\underline\delta})}=0=
(\Gamma^{\underline a})^{{\underline\alpha}({\underline\beta}}
(\Gamma_{\underline a})^{{\underline\gamma}{\underline\delta})}
$$

In the world manifold, the tangent Minkowski metrics
$\hat\eta_{{\hat a}{\hat b}}$ in d=3 and $\eta_{ab}$ in d=2 have
again signature such that  $\hat\eta_{00}=\eta_{00}=1$,
$(\hat a=0,1,2$ and $a=0,1$). We denote $\hat\sigma^{\hat a}$ and
$\sigma^a$ the real and symmetric ``$\Gamma$--matrices" in d=3 and in
d=2 respectively (2$\times$2 matrices in both cases). By choosing for the
Dirac and charge conjugation matrices in d=3 the representation

$$
\hat\gamma^0 = \tau_2 ;\quad \hat\gamma^1 = - \tau_1;\quad
\hat\gamma^2=i\tau_3 ;\quad C=-i\tau_2
$$

\noindent
we have

$$
\hat\sigma^0 = \sigma^0=1 ;
\hat\sigma^1=\sigma^1=\tau_3; \hat\sigma^2=\sigma^\perp= \tau_1
$$

\noindent
Moreover we define

$$
\sigma^\pm  = {1\over\sqrt{2}} (\sigma^0 \pm \sigma^1)
$$

\noindent
so that

$$
\sigma^+=\sigma_-= {\sqrt{2}\ 0 \choose 0\ 0};\quad
\sigma^- = \sigma_+ = {0\ 0\choose 0\ \sqrt{2}}
$$

\vfill\eject

\noindent
{\bf References}

\vskip 0.5truecm

\item{1]} D.P. Sorokin, V.I. Tkakch and D.V. Volkov,
Mod. Phys. Lett. {\bf A4} (1989) 901;
\item{} D.P. Sorokin, V.I. Tkach, D.V. Volkov and A.A. Zheltukhin,
Phys. Lett. {\bf 216B} (1989) 302.

\item{2]} D.P. Sorokin, Fortsch. Phys. {\bf 38} (1990) 302.
\item{} A.I. Gumenchuk and D.P. Sorokin, Sov. J. Nucl. Phys.
{\bf 51} (1990) 350;
\item{} V.A. Soroka, D.P. Sorokin, V.I. Tkach and D.V. Volkov,
Int. J. Mod. Phys. {\bf A7} (1992) 5977;
\item{} A.I. Pashnev and D.P. Sorokin, Class. Quantum. Grav.
{\bf 10} (1993) 625.

\item{3]} D.V. Volkov and A.A. Zheltukhin, Letters in Math. Phys.
{\bf 17} (1989) 141; Nucl. Phys. {\bf B335} (1990) 723.

\item{4]} N. Berkovits, Phys. Lett. {\bf 232B} (1989) 184; {\bf 241B}
(1990) 497; Nucl. Phys. {\bf B350} (1991) 193; {\bf B358} (1991) 169;
{\bf B395} (1993) 77.

\item{5]} M. Tonin, Phys. Lett. {\bf 266B} (1991) 312; Int. J. Mod. Phys.
{\bf A7} (1992) 6013;
S. Aoyama, P. Pasti and M. Tonin, Phys. Lett. {\bf 283B} (1992)
213; I. Bandos, D. Sorokin, M. Tonin and D. Volkov, Phys. Lett.
{\bf 319B} (1993) 445; D. Sorokin and M. Tonin, Phys. Lett.
\underbar{\bf 326} (1994) 84.

\item{6]} P. Howe and P. Townsend, Phys. Lett. {\bf 259B} (1991) 285.
\item{} F. Delduc and E. Sokatchev, Class. Quantum Grav. {\bf 9} (1991)
361;
\item{} A. Galperin and E. Sokatchev, Phys. Rev. {\bf D46} (1992) 714.

\item{7]} E.A. Ivanov and A.A. Kapustnikov, Phys. Lett. {\bf 267B}
(1991) 175;
\item{} F. Delduc, E. Ivanov and E. Sokatchev, Nucl. Phys. {\bf B384}
(1992) 334;
\item{} A.S. Galperin, P.S. Howe and K.S. Stelle, Nucl. Phys.
{\bf B368} (1992) 281;
\item{} F. Delduc, A. Galperin and E. Sokatchev, Nucl. Phys. {\bf B368}
(1992) 143.

\item{8]} J.P. Gauntlett, Phys. Lett. {\bf 272B} (1991) 25.

\item{9]} F. Delduc, A. Galperin, P. Howe and E. Sokatchev, Phys. Rev.
{\bf D47} (1992) 578.

\item{10]} V. Chikalov and A. Pashnev, Mod. Phys. Lett. {\bf A8} (1993)
285.

\item{11]} A. Galperin and E. Sokatchev,
Phys. Rev. {\bf D48} (1993), 4810.

\item{12]} P. Pasti and M. Tonin, Nucl. Phys. \underbar{\bf B418} (1994)
337.

\item{13]} I. Bandos, M. Cederwall, D. Sorokin and D. Volkov, preprint
ITP-94-10 G\"oteborg (1994).

\item{14]} E. Bergsoeff and E. Sezgin, preprint CTP TAMO-67/93,
hep--th/9312168

\item{15]} R. Penrose, J. Math. Phys. {\bf 8} (1967) 345;
R. Penrose and M.A.H. MacCallum, Phys. Rep. {\bf 6} (1972) 241
and refs. therein.
E. Witten, Nucl. Phys. {\bf B226} (1986) 245;
\item{} L.L. Chau and B. Milewski, Phys. Lett. {\bf 216B} 81989) 330;
\item{} L.L. Chau and C.S. Lim, Int. J. Mod. Phys. {\bf A4} (1989) 3819;
\item{} J.A. Shapiro and C.C. Taylor, Phys. Rep. {\bf 191} (1990) 221;
\item{} A. Ferber, Nucl. Phys. {\bf B132} (1978) 55;
\item{} T. Shirafuji, Progr. Theor. Phys. {\bf 70} (1983) 18;
\item{} A.K.H. Bengston et al., Phys. Rev. {\bf 36D} (1987) 1786;
\item{} I. Bengtsson and M. Cederwall, Nucl. Phys. {\bf B302} (1988) 81;
\item{} Y. Eisenberg and S. Solomon, Nucl. Phys. {\bf B309} (1988) 709;
\item{} M. Plyushchay, Mod. Phys. Lett. {\bf A4} (1989) 1827.
\item{} D.B. Fairlie and C.M. Monogue, Phys. Rev. \underbar{\bf D36} (1987)
475.

\item{16]}  E. Sokatchev, Class. Quantum Grav. {\bf 4} (1988) 237;
\item{} E. Nissimov, S. Pacheva and S. Solomon, Nucl. Phys.
{\bf B296} (1988) 492; {\bf B297} (1988) 349; Int. J. Mod. Phys.
{\bf A4} (1989) 737.
\item{} R.E. Kallosh and M.A. Rahmanov, Phys. Lett. {\bf 209B} (1988) 233;
{\bf 214B} (1988) 549;
\item{} I.A. Bandos, Sov. J. Nucl. Phys. {\bf 51} (1990) 906;
\item{} I.A. Bandos and A.A. Zheltukhin, Sov. Phys. JETP Lett.
{\bf 51} (1990) 547;
JETP Lett. {\bf 54} (1991) 421;
JETP Lett. {\bf 55} (1992) 81; Phys. Lett. {\bf 261B}
(1991) 245; Theor. Math. Phys. {\bf 88} (1991) 358; Fortsch. der Phys.
{\bf 41} (1993) 619; Phys. Lett. {\bf 288B} (1992) 77;
Sov. J. Nucl. Phys. {\bf 56} (1993) 198; Int. J. Mod. Phys. {\bf A8} (1993)
1081.

\item{17]} M.J. Duff, P.S. Howe, T. Inami and K.S. Stelle, Phys. Lett.
{\bf 191B} (1987) 70;
\item{} A. Ach\'ucarro, P. Kapusta and K.S. Stelle {\bf 232B} (1989)
302.

\item{[18]} J. Carr, S.J. Gates jr. and R. Oerter, Phys. Lett. {\bf 189B}
(1987) 68.

\item{19]} J.A. De Azc\`arraga, J.M. Izquierdo and P.K. Towsend,
Phys. Rev. {\bf D45} (1992) 3321;
\item{} P.K. Townsend, Phys. Lett. {\bf 277B} (1992) 285.
\item{} E. Bergshoeff, L.A.J. London and P.K. Townsend. Class. Quantum
Grav. {\bf 9} (1992) 2545.

\bye